# An Evidential Interpretation of the 1$^{st}$ and 2$^{nd}$ Laws of Thermodynamics

## VJ Vieland

The Research Institute at Nationwide Children's Hospital & The Ohio State University
575 Children's Crossroad, Columbus OH 43215
Phone: (614) 355-5651
email: veronica.vieland@nationwidechildrens.org

**Abstract** I argue here that both the 1$^{st}$ and 2$^{nd}$ laws of thermodynamics, generally understood to be quintessentially physical in nature, can be equally well described as being about certain types of information without the need to invoke physical manifestations for information. In particular, I show that the statistician's familiar likelihood principle is a general conservation principle on a par with the 1$^{st}$ law, and that likelihood itself involves a form of irrecoverable information loss that can be expressed in the form of (one version of) the 2$^{nd}$ law. Each of these principles involves a particular *type* of information, and requires its own form of bookkeeping to properly account for information accumulation. I illustrate both sets of books with a simple coin-tossing (binomial) experiment. In thermodynamics, absolute temperature T is the link that relates energy-based and entropy-based bookkeeping systems. I consider the information-based analogue of this link, denoted here as E, and show that E has a meaningful interpretation in its own right in connection with statistical inference. These results contribute to a growing body of theory at the intersection of thermodynamics, information theory and statistical inference, and suggest a novel framework in which E itself for the first time plays a starring role.

**Introduction**

As is well known, important connections exist between statistical mechanics and information theory. It has been widely recognized for some time that the concepts of entropy developed by Gibbs and others have, at least under some well circumscribed conditions, the same form as Shannon's [1] measure of average information, or Shannon entropy. The information-entropy connection has been further developed in connection with Gibbs' paradox [2] and Maxwell's demon [3] and elsewhere.

The link between physics and information theory can be made by treating information in its physical embodiment, that is, by recognizing that the extraction of information requires physical action, and then describing that action in familiar physical terms [4]. This is one way to reconcile the observation of identical bits of mathematics governing two topics – physics and information theory - that seem to be very distinct metaphysically. Jaynes [5] seemed to be getting at something more abstract, namely, that certain aspects of physics ought themselves to be understood in less physical terms. In particular, he argued that it is the probabilistic aspects of entropy, rather than physical aspects of matter, that give rise to the phenomena encapsulated by the second law of thermodynamics (henceforth simply called the 2$^{nd}$ Law). (See also [6,7].) More recently, Duncan and Semura [8,9] have argued that the most fundamental form of the 2$^{nd}$ Law is wholly information-based. They too, however, justify this statement of the 2$^{nd}$ Law in part by assuming physical embodiment for information loss.

Here I postulate the utility of decoupling Duncan & Semura's statement of the 2$^{nd}$ Law from its physical embodiment, and simultaneously extend this type of reasoning to consider thermodynamics as a whole from a purely informational (non-physical) point of view. I add to their statement of the 2$^{nd}$ Law an information-based version of the 1$^{st}$ Law of Thermodynamics (henceforth simply the 1$^{st}$ Law), and I develop a methodological framework for considering the dynamics of certain kinds of information flow adherent to these laws. Insofar as this new framework is coherent, this suggests that the dichotomy between what is physical and what exists purely in the realm of information is an unnecessary one from the start, not because information must have physical embodiment, but because as a matter of mathematics these two representations - physical and informational – are identical, at least over certain domains, without the need to posit physical existence for any of the constituent terms.

It is worth noting up front that in other information-theoretic treatments involving elements of thermodynamics, (the analogue of) temperature itself plays little or no role. E.g., in the work of authors like Caticha [6] and Bialek [10] employing maximum entropy as an inferential device, the Lagrange multiplier as it appears in the Boltzmann distribution is given no particular interpretation beyond its role as a calculation device. This is in no way a criticism of this very interesting body of work, but any attempt to recapitulate thermodynamics *as a whole* on a purely information-based footing must include a salient role for an informational equivalent to absolute temperature T: after all, it is T that puts the "thermo" in "thermodynamics."

In a previous paper [11], my coauthors and I set out to derive an absolute scale for the measurement of statistical evidence. We posited that this could be accomplished by harnessing the mathematical underpinnings of T, and constructed a proof of principle for a particularly simple (binomial) statistical model. The current paper traverses some of the same terrain, however, with a very different orientation: while [11] started with basic physical quantities and relied heavily on analogy to link these to informational counterparts, here I build up an "information dynamics" framework in purely mathematical terms. Again, however, in the current treatment the concept of statistical evidence plays a central role, mathematically equivalent to the central role of temperature in thermodynamics.

**Results**

Because the notion of statistical evidence ends up playing so central a role, I begin with (i) a brief account of the sense in which the term "evidence" is invoked throughout the paper. I then develop the general framework in three steps. In (ii) I derive a formal (purely mathematical) connection between the familiar likelihood principle in statistics and the 1$^{st}$ Law. In (iii) I similarly relate a principle concerning information loss inherent in the use of likelihood to (a form of) the 2$^{nd}$ Law. Then in (iv) I consider the link between these two principles. If we assume that this link has the same form in the new informational framework as it does in thermodynamics, then we arrive at the very surprising conclusion that thermodynamic T has a purely information-based interpretation in addition to its interpretation as temperature, viz., as the strength of statistical evidence, in a sense to be made clear.

Throughout, I develop the theory with respect to a single, and particularly simple, binomial statistical set up. In part this is simply to facilitate the exposition through concrete examples and calculations. But in part, aspects of this set up are fundamental to the framework as currently developed, a point I return to in the **Discussion**. Details of calculations used to generate the figures are given in **Methods**, below.

(i) ***Properties of evidence*** In much of the literature relating information theory, thermodynamics and statistical inference, the term "evidence" is used casually and without explicit definition. This is true of work that is otherwise extremely rigorous, thus it reflects the fact that evidence *per se* is simply not being considered as a central concept in this literature. When the term is defined, it is sometimes treated as synonymous with "data," or in other instances used to refer to the (marginal) probability of the data (the normalizing constant in the Boltzmann distribution). Here I am using the term in neither of those senses, but in a sense that is more closely related to some statistical literature (see, e.g., [12]). Therefore, I begin with an account of evidence in the sense in which it plays a role in the current framework. I do so by considering the behavior of evidence informally, in a simple setting in which our intuitions are clear. This allows us to enumerate key properties of what we *mean* by evidence, properties that any formal treatment of evidence would need to recapitulate.

Consider a series of *n* independent coin tosses of which *x* land heads (H). And consider the two hypotheses "coin is biased towards tails ($\bar{H}$)" versus "coin is fair" (i.e., the coin lands $\bar{H}$ and H with equal probabilities; I use the notation $\bar{H}$ to designate tails, since the more obvious choice "T" is already being used for absolute temperature). What follows are some thought experiments appealing to the reader's statistical intuition.

Suppose that on repeated tosses the coin consistently lands H (approximately) 5% of the time. Clearly as the number of tosses *n* increases, if *x/n* remains around 5%, the evidence that the coin is biased *increases*. Now consider a fixed number of tosses *n*, but allow *x/n* to increase from 5% upwards. Here the evidence in favor of bias clearly *decreases* as the proportion *x/n* increases in the direction of ½. These two features together entail a third, somewhat more abstract property. Suppose now that we hold the *evidence* constant (without defining it or saying how to calculate it). If *x/n* starts at 0.05, what must happen as *n* increases, in order to prevent the evidence itself from increasing? If the previous two properties hold, then it follows that in this scenario, *x/n* must increase; otherwise the evidence would increase as *n* increases rather than remaining constant. Thus the three quantities *n*, *x*, and evidence *e*, enter into an equation of state, in which holding any one of the three constant while allowing a second to change necessitates a compensatory change in the third. Here *e* itself is simply *defined* as the third fundamental entity in the set.

I note two additional and very important properties of *e* up front. First, for given *n*, as *x/n* increases from some very small value towards ½, the evidence for "bias" at first decreases, but the evidence remains in favor of bias up to some particular value of *x/n*. Beyond that value, however, as *x/n* continues to approach ½ from the left, the evidence switches to favoring "no bias." I refer to the value of *x/n* at which this switch occurs as the "transition point" (TrP). To the right of the TrP, as *x/n* continues to increase beyond this point up to $x = n/2$, the evidence for "no bias" *increases*. Figure 1 illustrates this pattern. (Statistical intuition also suggests that TrP(E) itself ought to move towards $x = n/2$ as *n* increases, so that even a small deviation from $x = n/2$ would correspond to evidence against $\theta = ½$ for very large *n*.) A mirror-image set of properties occur as *x/n* increases to the right of $\theta = ½$. In what follows, I will focus on the one-sided hypothesis contrast "coin is biased towards $\bar{H}$" ($\theta < ½$) versus "coin is fair," so that we need not consider $\theta > ½$.

Finally, note that the same quantity of new data (*n*, *x*) seems to have a *smaller* impact on our intuitive assessment of the evidence the *larger* is the starting value of *n*, or equivalently, the stronger the evidence is before consideration of the new data. E.g., suppose we have already tossed the coin 1,000 times and observed $x = 0$. We will all agree that this is overwhelming evidence of a biased coin. If we toss the coin an additional 5 times and observe 0 H, the new data (5, 0) change the strength of the evidence hardly at all. However, had we started with just 2 tosses with 0 H, then added the same new data (5, 0), the evidence in favor of bias would have changed quite a bit. The initial set (2, 0) gives only very slight evidence for bias, but by the time we have observed (7, 0), we would be far more suspicious that the coin is not fair. Despite the purely qualitative nature of this example, it illustrates the key point that evidence is not equivalent to or even inherent in data *per se*: the evidence conveyed by new data depends on the context in which they are observed.

Table 1 summarizes these key properties of evidence. They play no particular role in §(ii) and §(iii) below, but they become critically important in §(iv). In the mean time, they serve to motivate the use of the qualifier "evidential" in some of the following text.

**(ii) *The Likelihood Principle and the 1st Law of Thermodynamics*** The familiar likelihood principle in statistics states that all of the information conveyed by a set of data regarding a parameter (vector) is contained in the likelihood (see, e.g., [12]). Because below I will need to distinguish this type of information from another, I refer to the information conveyed by the likelihood as "evidential information." The other type of information, which I'll call "combinatoric information," is introduced in §(iii).

To be concrete, again consider an experiment comprising $n$ independent coin tosses with $x$ = the number of H, and P(H) = $\theta$. The likelihood is defined as

$$L(\theta|n,x) \propto P_n(x|\theta) = k\theta^x(1-\theta)^{(n-x)}, \qquad \text{Eq 1}$$

where k is an arbitrary scaling constant. We are interested in comparing the hypotheses $\theta < \frac{1}{2}$ (coin is biased toward $\overline{H}$) vs. $\theta = \frac{1}{2}$ (coin is fair). Then the likelihood principle further implies that all of the information distinguishing these hypothesis for given data ($n$, $x$) is contained in the likelihood ratio (LR) between these two hypotheses:

$$LR(\theta|n,x) = \frac{k\theta^x(1-\theta)^{(n-x)}}{k\left(\frac{1}{2}\right)^n} = 2^n\theta^x(1-\theta)^{n-x}. \qquad \text{Eq 2}$$

I will refer to this extrapolation of the likelihood principle from the likelihood itself to the LR as the extended likelihood principle; again, see, e.g., [12]. Note that Eq 2 is frequently treated in the statistical literature as expressing the *evidence* for (or against) "coin is biased" vs. "coin is fair." I return to this interpretation of the LR in the **Discussion**.

Now consider two sets of data. Let $D_1$ comprise $n_1$ tosses of which $x_1$ have landed H. Denote the graph of the corresponding LR($\theta|n,x$), plotted over $\theta$ = [0, ½] on the x-axis, as LR(A). We are interested in the effects on LR(A) of a second data set, $D_2 = (n_2, x_2)$. Let the corresponding graph, considering both $D_1$ and $D_2$, be LR(B). The consideration of $D_2$ results in a transformation of the graph from its initial state A to a new state B. Even prior to considering the nature or mechanism of that transformation, one requirement is that it must appropriately reflect the effects of the new data and nothing but the effects of the new data. Otherwise, the transformation would lead to a violation of the extended likelihood principle. Thus we will need to characterize transformations of the LR graph in terms of an underlying state variable, which I'll denote by U. Requiring U to be a state variable means that, for any given set of data $(n, x)$, U must depend only on the LR for given data, and not, for example, on anything related to the history of data collection.

I now seek to formally characterize the change in U from state A to state B, or $\Delta U$, corresponding to the change $\Delta LR$. The binomial LR graph for given $n, x$ can be uniquely specified in terms of two quantities, but there is leeway regarding which two we choose. For example, we could use $(n, x)$ itself, but this turns out to be not particularly revealing in the current context. Alternatively, we can uniquely specify the LR in terms of properties of the LR graph. Here I use the area under the LR curve, denoted by V, and a second quantity denoted by P. P is chosen such that the pair (V, P) uniquely determines a particular LR graph, and such that for given U, P is inversely proportional to V. (The existence of a quantity P fulfilling these conditions in the binomial case is shown in [11]. Note that throughout I treat V and P as continuous rather than discrete; see **Methods**.) For the moment there is no special significance to the choice of variable names for these quantities. However, as a mnemonic device, the reader may consider them as counterparts of volume (V) and pressure (P), respectively, while U is the counterpart of internal energy. Note that the proportionality requirement for P enforces an "ideal gas" relationship to V.

To consider a concrete example of the bookkeeping involved in quantifying $\Delta U$, suppose that $D_1 = (n,x) = (4,1)$ and $D_2 = (2,1)$. Thus LR(A) corresponds to (4, 1) while LR(B) corresponds to (6, 2). There are two possibilities regarding how we may have gotten from LR(A) to LR(B): either the 5[th] toss landed H and the 6[th] toss landed $\bar{H}$, or, the 5[th] toss landed $\bar{H}$ and the 6[th] toss landed H. That is, the transformation from LR(A) to LR(B) could go through the intermediate states (5,1) or (5,2). Call these two possibilities Path 1 and Path 2, respectively. Figure 2 shows the two paths plotted both in terms of the LR and on a classical PV diagram.

Since both paths begin and end with identical LR graphs, our state variable U must change by the same amount in both cases. Thus we would like to find a function of V, P such that dU(V, P) is a perfect differential. By stipulation, $VP \propto U$, so that $dU = c(VdP + PdV)$ has the correct form, where c is the constant of proportionality. Define the quantities W = PdV and Q = cVdP + (c+1)PdV. Note that both W and Q are readily verified to be path-dependent, that is, their values will be different for the two transformations. Then we have

$$dU = c(VdP + PdV) = (cVdP + [c+1]PdV) - PdV = Q - W. \qquad \text{Eq 3}$$

Beyond the choice of variable names, nothing whatsoever in the derivation makes any reference to physical quantities. Here W is simply one aspect of the transformation from LR(A) to LR(B), and Q is the compensatory quantity required to express changes in U (a state variable) in terms of W (a path-dependent variable). Then Eq 3 tells us that the amount of evidential information "received" by the LR as a result of new data is (Q-W).

Armed with this new notation, we now have a concise way to express the principle that all of the evidential information is contained in the LR: ΔU = Q - W, or equivalently, Q = ΔU + W. The extended likelihood principle can therefore be reformulated to state that *the variation in evidential information of a system during any transformation* (ΔU) *is equal to the amount of evidential information that the system receives from new data* (Q − W). (This statement is paraphrased from Fermi's general statement of the 1st Law [13], p. 11.) Thus the extended likelihood principle, which requires that transformations of the LR graph (up to allowable scale transformations) reflect all and only changes in the data, turns out to be a fundamental conservation principle on a par with the 1st Law.

**(iii) *Information loss and the 2nd Law of Thermodynamics*** I continue with the coin-tossing example. But whereas §(ii) dealt with what I called "evidential information," I now consider the other *type* of information, or "combinatoric information." In a sequence of *n* independent tosses of a single coin, each toss can land either H or $\bar{H}$. Hence if we know the actual sequence of tosses we have

$$I_{SEQ} = \ln 2^n = n \ln 2 \qquad \text{Eq 4}$$

units of information, corresponding to the logarithm of the number of possible sequences. (For present purposes the base of the logarithm, which further defines the units of information, is unimportant. Note that Eq 4 is more commonly thought of as the amount of *un*certainty or entropy associated with having no knowledge of the sequence; but it will be more convenient here to consider this as the information corresponding to complete knowledge of the sequence. See also [1].) Obviously $I_{SEQ}$ increases linearly in *n*. This makes sense: all other things being equal, the amount of information goes up with the number of tosses.

Now suppose that, rather than knowing the full sequence of H and $\bar{H}$, we know only the number *x* of H (hence the number (*n-x*) of $\bar{H}$). In this case, we know only that the actual sequence was one of $\binom{n}{x}$ possibilities, and the amount of information we have is

$$I_{(n,x)} = ln \binom{n}{x} < I_{SEQ}. \qquad \text{Eq 5}$$

The change in information in going from $I_{SEQ}$ to $I_{(n,x)}$, written with a negative sign to indicate lost information, is

$$-\Delta I \triangleq I_{SEQ} - I_{(n,x)}. \qquad \text{Eq 6}$$

Clearly the behavior of $-\Delta I$ is a function of both *x* and *n*. It is not guaranteed to increase as *n* increases, because the inherent increase in information as *n* increases is offset by a corresponding increase in the amount of information *lost* in going from $I_{SEQ}$ to $I_{(n,x)}$, which is further mediated by the value of *x/n*. Thus the underlying dynamics here requires an accounting system that simultaneously takes into account changes in *n* and changes in *x*.

I would argue that this particular type of combinatoric information loss is a ubiquitous feature of abstract reasoning, or the generalization from "raw" information to explanatory principles. For instance, suppose that what we are really interested in is information *about* the probability θ that the coin lands H. In order to extract this information *about* θ, we need to rewrite the sequence of H and $\bar{H}$ as an expression in terms of θ. This entails a compression of the original sequence into the expression $\theta^x(1-\theta)^{(n-x)}$, or in logarithmic terms,

$$g(\theta) = x\, ln\, \theta + (n-x)\, ln\, (1-\theta) = ln\, L(\theta \mid n,x), \qquad \text{Eq 7}$$

which is simply the ordinary ln likelihood for given data, i.e., the ln of Eq 1. (Thus information *about* θ appears to return us to the "evidential information" conveyed by the likelihood as discussed in the previous section, while the current section deals with some other type of information; see e.g., [14] for a related distinction.) But the combinatoric information associated with Eq 7 is $I_{(n,x)} < I_{SEQ}$, because once the sequence of H, H̄ is reduced to the form of Eq 4, the full sequence can no longer be reconstructed. The lost combinatoric information has been permanently erased.

This illustrates a very general principle: data reduction for purposes of gleaning information *about* underlying parameters always comes at a price. *Any process that extracts information regarding an underlying parameter (vector) from a set of data entails irrecoverable loss of information, or the permanent erasure of some of the information associated with the full data prior to compression.* This is in essence the form of the 2$^{nd}$ Law proposed by [8], to which I have added a specific context in which information is routinely erased, a context that makes it clear that we can consider irrevocable information loss in accordance with the 2$^{nd}$ Law without having to postulate physical existence for the information or for the information lost through data reduction.

**(iv)** ***The information-theoretic analogue of T*** Finally, I consider the connection between transformations of the LR graph characterized in terms of changes in evidential information (ΔU from §(ii)) and changes in combinatoric information (ΔI from §(iii)). A picture emerges of a complex information-dynamic system. (Thinking in terms of dynamics seems useful here, even if transformations are not considered as functions of time.) All other things being equal, the more data we have, the more evidential information we have. But this information gain is mediated by a corresponding increase in the amount of combinatorial information erased in the process, which is a function (in the binomial case) of both *n* and *x*. At the same time, while the amount of evidential information is increasing, evidence itself is not necessarily increasing, because it again is a function of both *n* and *x*. E.g., evidence in favor of the hypothesis that the coin is biased could go down going, say, from (*n* = 4, *x* = 0) to (*n* = 8, *x* = 4), despite the doubling of the sample size. Both evidential information and combinatoric information are in play, and this suggests that we need a way to link the bookkeeping expressed in terms of combinatoric information with the bookkeeping expressed in terms of evidential information. The only remaining step, then, is to articulate this link.

Following [8], I do this by introducing a quantity E, the analogue of thermodynamic temperature T, as the proportionality factor linking the two sets of books. That is, I postulate that E relates U, W and ΔI, as these quantities are defined above in terms of information, through the same equation used to relate T to the corresponding quantities in thermodynamics. Thus following [8] but writing E instead of T we have,

$$E = \frac{\Delta U + W}{-k\Delta I} = \frac{Q}{-k\Delta I} \qquad \text{Eq 8}$$

where *k* is a constant (not necessarily equal to Boltzmann's constant). Eq 8 relates the incoming evidential information transferred in the form of Q to the net loss of combinatoric information.

But what reason do we have for thinking that the relationship expressed between the two sides of Eq 8 has any useful meaning, given that we are not interpreting W as mechanical work or Q as physical heat? In the current context, E plays a purely abstract role, as the link relating evidential and combinatoric information. Of course, Kelvin's derivation of T was also quite abstract, arguably a matter more of calculus than physics, and historically predating our understanding of thermal energy in the terms of statistical mechanics. (For a fascinating account of exactly how difficult it was, and remains to this day, to directly relate T to actual physical phenomena, see [15], especially Chapter 4.) Notwithstanding, T itself has a critically important physical interpretation in the theory of *thermo*dynamics. It behooves us, therefore, to seek a corresponding information-based interpretation of E.

Figure 3 illustrates the behavior of E, as defined by Eq 8, as a function of (*n*, *x*). It is readily confirmed that in its behavior, E recapitulates the principal properties ascribed to evidence in §(i) above (per Table 1). Thus E can be understood to be the *evidence* measured on an absolute scale. (That is, E is the analogue of T, while the quantity *e* in §(i) above is the analogue of thermodynamic *t*, or temperature measured on an arbitrary scale. See also [11] for additional detail on E as a formal measure of evidence.) It would appear, then, that a complete alignment of thermodynamics with this new purely information-based framework must entail both the new interpretations given above for the 1st and 2nd Laws of thermodynamics, as well as pride of place for this new quantity E, which appears in the end as the thing the new theory is actually *about*: the *evidence*.

I want to stress that in terms of the theory as it has been derived here, the relationship between E and evidence is an empirical discovery. E was introduced as the link connecting two types of information bookkeeping, assuming a system governed by purely information-based versions of the 1st and 2nd Laws, and under the postulate that E would have the same mathematical form as its analog T in thermodynamics. E need not have turned out to have any recognizable behavior or meaningful interpretation. The fact that it turns out to have an interpretation as evidence, a concept so seemingly fundamental to inference, strongly supports the idea that thermodynamics is serving here as more than mere analogy. The mathematical underpinnings of thermodynamics appear to relate to the "dynamics" of information flow just as directly as they relate to heat flow.

**Discussion**

Here I have derived a conceptual framework in which the 1$^{st}$ Law, the 2$^{nd}$ Law, and even T itself (here renamed E) play their usual thermodynamic roles without the need to posit physical existence for any of the underlying quantities. This framework sets the stage for a new look at statistical inference and, more generally, any type of mathematical modeling the aim of which is to assess evidence for or against various models or hypotheses. The framework presented in this paper is an outgrowth of work towards a measure of statistical (or more generally, mathematical) evidence E that is on an absolute scale, in the same sense in which Kelvin's T is on an absolute scale [16-18]. And indeed, the derivation of E as described above and in [11] may be the most important practical implication of this work. But there are also other interesting aspects of the formalism as it relates both to thermodynamics and the new framework of evidentiodynamics (or perhaps simply "evidentialism" for lack of a better word).

First, as we argued in [11], there is good reason to think that statistical systems are always and instantaneously in their maximum entropy states: that is, there seems no reason whatsoever to postulate temporal processes corresponding to transformations of the LR graph; and no obvious reason, at least so far, to consider non-equilibrium conditions on the evidential side. Yet we appear to require an evidential version of the 2$^{nd}$ Law, and moreover, to have a cogent statement of that law that neither refers to time's arrow nor postulates some inevitable drift of systems towards maximum entropy states. This seems consistent with the claim in [8] that an entirely information-based articulation of the 2$^{nd}$ Law for thermodynamics itself is more fundamental than other contenders. (See also [19] for related discussion.) What I have added to this is an understanding of the 2$^{nd}$ Law in terms of a particular type of information loss, namely, combinatoric information loss through abstraction of information *about* something, or what we could simply think of as the process of (data analytic) measurement itself.

Note also that in [11] we defined the evidential analogue of thermodynamic entropy as *relative* entropy, per Eq 10 above, rather than Shannon entropy. And in fact were we to use the Shannon entropy itself, the resulting quantity E would not behave like evidence, at least not without making other adjustments to the underlying system. Since in the case of the binomial distribution, evaluating the Shannon entropy at $\theta = \hat{\theta}$ corresponds to the maximum entropy constrained by the observed ($n$, $x$), while evaluating it at $\theta = \frac{1}{2}$ corresponds to the unconstrained maximum entropy, it is perhaps not surprising that this quantity should play a key role in a thermodynamically-based description of the system. This appears to be related to the view sometimes stressed by Jaynes [20] and some others that thermodynamic entropy is related to but not the same as Shannon entropy.

A second point of comparison relates to what we are most interested in when we view the framework in evidential rather than thermodynamic terms. In [11] we showed how we could run Carnot cycles for purely evidential systems, in order to derive a corresponding version of evidential efficiency and thus formal derivation of the analogue E of Kelvin's T. Historically, Carnot's work took place in connection with the search for maximally efficient heat engines. The point of such engines (expressed in modern terms) was to convert heat into work, that is, work was the desired outcome of running the engine. In this context, the amount of input energy that cannot be converted to work is only of interest insofar as we are able to minimize it. On the evidential side, however, the situation is reversed. Viewed in terms of an interest in the change in evidence as new data are considered, it is the "work" W itself that represents lost "energy," or information in the new data that gets dissipated rather than transformed into a change in the "internal energy" U (hence, evidence E) of the system. This perhaps also suggests an interesting new way to think about mechanical work itself in terms of information.

A third important observation is that bridging the gap between the theory as it applies to physics and as it applies to non-physical information dynamics appears to require understanding these dynamics from a quintessentially thermodynamic perspective, rather than in terms of statistical mechanics. This is counterintuitive, and to my knowledge, perhaps even unique among discussions of the connection between physical entropy and information entropy. It seems to me that the deepest connections between physics and information dynamics are to be found primarily at the "macroscopic" level. Indeed, it is far less clear to me how to make mathematical sense of non-physical analogues of particles and their motions. For instance, the individual data points in the binomial system do not seem to be analogues of individual particles in physical systems; see [11] for discussion of $n$ (together with $x$) as an index of "energy" rather than as a measure of data quantity (the analogue of the number of particles). Thus despite the temptation to relate the theory of statistics to the statistical view of mechanics, thermodynamic theory itself, without recourse to descriptions of microstates, seems the more fundamental view of things. This seems consonant with the views of Callen [21] on the universality of the theory of thermodynamics, which now appears fundamental to an even broader swath of science than he had envisioned. Possibly statistical mechanics describes just one of multiple possible microscopic systems that all give rise to the same macroscopic behavior embodied by thermodynamics in its most abstract form.

Still unclear, however, is the nature of the relationship between the coin tossing set up (under the one-sided hypothesis contrast) and thermodynamic theory. In order to represent thermodynamics as a whole in information-theoretic terms, physicists would need an extension beyond the ideal gas (independence) model to incorporate more realistic and complex types of physical behavior; while for statistical purposes, we need to know how to apply the framework to probability distributions beyond the binomial. But have we developed the theory only far enough to *apply* it to binomial systems, or, have we discovered that there is some fundamental relationship between the binomial model and elementary thermodynamics as a fundamentally macroscopic theory? (That is, remembering that the binomial distribution in this context is not describing the behavior of particles.) If the latter, then how would we *use* the theory in application to more complex systems?

The fourth and final point concerns implications of this framework for statistical treatments of evidence. As noted above, the LR itself is sometimes interpreted as representing the statistical evidence. In many contexts, this interpretation is assigned to the $\max \text{LR} = \text{LR}(\hat{\theta})$ (to continue for notational simplicity with our binomial example having a single free parameter in the numerator of the LR and none in the denominator); in others the evidence is related to the Bayesian quantity $\int \text{LR}(\theta) \pi(\theta) \, d\theta$, where $\pi(\theta)$ is the prior distribution. In the current framework, however, $\text{LR}(\hat{\theta})$ would appear to be better understood as a measure of combinatoric information loss or relative entropy (see **Appendix**), while $\int \text{LR}(\theta) \pi(\theta) \, d\theta$ violates the extended likelihood principle by introducing information through $\pi(\theta)$ beyond what is conveyed by the data alone. (Note too that $\int \text{LR}(\theta) \, d\theta$ plays the role of V, not evidence; see **Methods**.) Thus the new framework is difficult to reconcile directly with other established accounts of the foundations of statistical inference. This is perhaps not so surprising, however, since it also differs from those accounts in its foundational emphasis on evidence and measurement [11], rather than starting from the more familiar statistical objectives of parameter estimation, hypothesis testing, quantification of statistical error rates, and/or the rational rank ordering of beliefs.

But the underlying theory is far from complete, and I mention here some important areas for future work. In the first place, some aspects of the theory clearly do need to be extended beyond the independence ("ideal gas") model. For instance, the notion of combinatoric information as deployed here probably needs to be generalized, since combinatorics *per se* do not play the same role in continuous probability distributions or some other discrete distributions. Nevertheless, as soon as we compress a full set of observations into, say, the sample average of a continuously distributed random variable, we have irrevocably lost some of the information associated with the original data, and the exact set of original values cannot be recovered from knowledge of the average alone. Thus the processes considered here appear quite general, although some of the formal apparatus developed above may be specific to the multinomial class of likelihoods. And as noted above, we still lack a template for applying the theory to non-binomial inference problems.

Moreover, it is not clear whether the current theory as it stands succeeds in solving a fundamental problem (what Callen [21], p. 26, considers to be *the* fundamental problem of thermodynamics), namely, how to determine the new state of a composite system once "barriers" between constituent subsystems have been removed. Restated in evidential terms, this is the problem of how to determine the new evidence upon concatenation of (constituent) data (sub)sets. The difficulty is that, in a thermodynamic system, energy will always flow from the warmer to the cooler body, and the reasoning in [8] and [10] shows that the information-based version of the 2$^{nd}$ Law maintains the prohibition against "energy" flow in the wrong direction. Evidence, however, does not on the face of it appear to follow this template (at least not obviously so), since whether new data increase or decrease the evidence for a given hypothesis seems intuitively as if it should depend only on whether the new data themselves represent evidence for the hypothesis, and not merely on whether they convey more or less evidence for the hypothesis relative to the current data. E.g., noticing the slight scent in the air corresponding to weak evidence that it is about to rain, after seeing the dark storm clouds corresponding to strong evidence that it is about to rain, increases rather than decreases the evidence for rain. See [16-18] for more detailed discussion of this issue. Reconciling the requirements of the 2$^{nd}$ Law with this pattern of behavior for evidence may be in part a matter of designing our measurement "apparatus;" e.g., perhaps measuring E through fixed volume, or zero "work," transformations. But it will almost certainly require a deeper understanding of these systems on the evidential side. (This topic relates to Dempster-Shafer theory in computer science; see, e.g., [23].) See also [11] for discussion of the role of apparent "phase transitions" in evidential systems as evidence moves from favoring the numerator of the LR to favoring the denominator, which may also factor into correct handling of the evidence concatenation problem. These evidential phase transitions thus also require further attention.

Finally, to end on a somewhat whimsical note, from this new perspective it might appear to be T – rather than E – that is the more mysterious quantity. Had evidentiodynamics preceded thermodynamics historically, we would undoubtedly be resistant to the notion that the proportionality factor connecting two forms of information bookkeeping could have purely *physical* existence, let alone existence as the basis for what we feel when we experience changes in temperature. That is indeed remarkable.

**Methods**

Numerical calculations used in generating the figures were calculated as described in detail in [11]. Briefly, we begin with the LR (Eq 2 above), and two key features of the LR graph. First, we define the area under the LR as

$$V = \int LR(\theta)d\theta = \int 2^n \theta^x (1-\theta)^{(n-x)} d\theta. \qquad \text{Eq 9}$$

Here the integral is taken over the interval θ = [0,..,½], reflecting the one-sided hypothesis contrast considered above. Second, we treat the maximum ln LR as a form of entropy, or evidential entropy, denoted $S_E$ (see the **Appendix** for additional details). From there we simply follow the mathematics of thermodynamics for ideal gases, setting $U = C_V E$, for $C_V$ a constant. This entails defining a quantity P through the relationship $PV \propto RE$, for R another constant, and assuming that the system follows the 1[st] and 2[nd] Laws. Following [13] (p. 147) this gives us

$$S_E = C_V \ln E + R \ln V. \qquad \text{Eq 10}$$

For purposes of these calculations, we set $C_V$ and R somewhat arbitrarily to 1.5 and 1 respectively. (Note that changing $C_V$ affects the size of the units of E. Note also that this sets the constant in Eq 3 to c = $R/C_V$. See [11] for additional discussion regarding the constants.) This gives

$$E = \frac{\exp\{S_E/C_V\}}{V^{R/C_V}}. \qquad \text{Eq 11}$$

Note that P can then be calculated from E and V as

$$P = \frac{RE}{V}. \qquad \text{Eq 12}$$

All calculations were done in MATLAB; numerical integrals were computed assuming continuous linear changes in *n* and *x*.

**Acknowledgments** This paper was written in the context of ongoing collaboration with J Das, SE Hodge and S-C Seok, all of whom have influenced development of the theory and provided valuable commentary on drafts of this paper; S-C Seok performed the calculations. I am also indebted to Hasok Chang and Gunter Wagner for their willingness to invest effort in following this work and for their many helpful critical suggestions, which have propelled the project forward. Finally, I thank the editor and the anonymous reviewers for their willingness to give this odd paper the benefit of the doubt. Their suggestions led to a substantially improved manuscript.

**Appendix:** *Relationships among likelihood, Shannon entropy, and evidential entropy $S_E$*

The general form of Shannon entropy for a discrete probability distribution is

$$S = -\sum_{x=0}^{n} f(x) \ln f(x), \qquad \text{Eq 13}$$

where $f(x)$ is the probability of $x$. To see the connection to the ln likelihood (Eq 7), rather than starting with the usual form of the binomial probability distribution, consider the complete set of ordered sequences of H and $\bar{H}$. E.g., with n = 3, there are $i = 2^3 = 8$ such sequences: $\bar{H}\bar{H}\bar{H}$, $\bar{H}\bar{H}H$, $\bar{H}H\bar{H}$, $H\bar{H}\bar{H}$, $\bar{H}HH$, $H\bar{H}H$, $HH\bar{H}$, $HHH$. For each sequence $i$, the probability $f(i) = \theta^x(1-\theta)^{n-x}$, where $x$ is the number of H in the sequence as above. Rewriting the summation over $x$ in Eq 13 as a summation over $i$ and then collecting like terms, we have

$$\begin{aligned}
-S &= \sum_{i=1}^{2^n}[f(i) \ln f(i)] \\
&= \sum_{i=1}^{2^n} \theta^x(1-\theta)^{n-x} \times \ln[\theta^x(1-\theta)^{n-x}] \\
&= \sum_{x=0}^{n} \binom{n}{x} \theta^x(1-\theta)^{n-x} \times \ln[\theta^x(1-\theta)^{n-x}] \\
&= E_x[x \ln \theta + (n-x)\ln(1-\theta)] = E_x[\ln L(\theta \,|\, n, x)], \qquad \text{Eq 14}
\end{aligned}$$

i.e., the expected value of Eq 7.

Maximizing Eq 7 yields the familiar maximum likelihood estimate (m.l.e.) $\hat{\theta} = x/n$, and evaluating Eq 7 at the m.l.e. reveals an (approximate) equivalence with the combinatoric information expressed in Eq 5 (using Stirling's approximation).:

$$\ln L(\hat{\theta} \,|\, n, x) = x \ln x + (n-x)\ln(n-x) - n \ln n \sim \ln \binom{n}{x} \qquad \text{Eq 15}$$

Evaluating Eq 2 at the m.l.e. of θ in the numerator and taking the ln yields

$$\begin{aligned}
\ln LR(\hat{\theta}) &= \ln \frac{\hat{\theta}^x(1-\hat{\theta})^{n-x}}{1/2^n} = \ln \frac{\left(\frac{x}{n}\right)^x \left(\frac{n-x}{n}\right)^{n-x}}{1/2^n} \\
&= x \ln x + (n-x)\ln(n-x) - n \ln n + n \ln 2. \qquad \text{Eq 16}
\end{aligned}$$

Now if we also evaluate Eq 13 at the m.l.e., which is also the value of θ that maximizes the Shannon entropy, and assume "perfect" data, or data occurring in their exact expected proportions (as would occur for very large values of *n*), then we have

$$-S(\hat{\theta}) = x \ln x + (n-x) \ln(n-x) - n \ln n, \qquad \text{Eq 17}$$

so that

$$\ln LR(\hat{\theta}) = -S(\hat{\theta}) + n \ln 2 = -\Delta I. \qquad \text{Eq 18}$$

Thus while max ln LR(θ) is frequently taken to be a measure of statistical *evidence*, Eq 18 suggests that it would be more appropriately understood as a measure of relative combinatoric *information*. (See also, e.g., [24, 25, 26] for other frameworks in which the LR is considered in information-theoretic terms.)

*arXiv September 2013*17. Vieland VJ (2011) Where's the Evidence? Human Heredity 71: 59-66.
18. Vieland VJ, Hodge SE (2011) Measurement of Evidence and Evidence of Measurement (Invited Commentary). Stat App Genet and Molec Biol 10: Article 35.
19. Uffink J (2001) Bluff your way in the second law of thermodynamics. http://arxiv.org/abs/cond-mat/0005327
20. Jaynes ET (2003) Probability Theory: The Logic of Science. New York: Cambridge University Press.
21. Callen HB (1985) Thermodynamics and an Introduction to Thermostatistics, 2nd Ed. New York John Wiley & Sons.
22. Vieland VJ, Huang Y, Seok SC, Burian J, Catalyurek U, et al. (2011) KELVIN: a software package for rigorous measurement of statistical evidence in human genetics. Human Heredity 72: 276-288.
23. Shafer G (1976) A mathematical theory of evidence. Princeton and London: Princeton UP.
24. Kullback S (1997) Information Theory and Statistics. New York: Dover.
25. Zellner A (1988) Optimal information processing and Bayes's theorem. Am Statistician 42: 278-280.
26. Soofi ES (2000) Principal Information Theoretic Approaches. J Amer Statist Assoc 95: 1349-1353.
18

**Table 1** Four defining properties of evidence (*e*)

| | Left of TrP (Evidence for bias) | | | Right of TrP (Evidence against bias) | | |
|---|---|---|---|---|---|---|
| **Property** | *n* | *x/n* | *e* | *n* | *x/n* | *e* |
| **P1** | ↑ | = | ↗ | ↑ | = | ↗ |
| **P2** | = | ↑ | ↘ | = | ↑ | ↗ |
| **P3** | ↑ | ↗ | = | ↑ | ↘ | = |
| **P4** | Δ*e* with new data depends on previous data | | | | | |

Patterns are shown corresponding to the thought experiments described in the text, in which two of the three quantities (*n*, *x/n*, *e*) follow a stipulated pattern and our intuition regarding the behavior of the third quantity is recorded. For P1-P3, ↑ indicates that the given quantity is stipulated as increasing; = indicates that the quantity is stipulated as constant; and ↗ and ↘ indicate the direction of change (increase or decrease, respectively) of the third quantity.

**Figure 1** Schematic illustration of the behavior of evidence *e* (blue curve) around the transition point TrP, as a function of increasing *x*/*n* for fixed *n*.

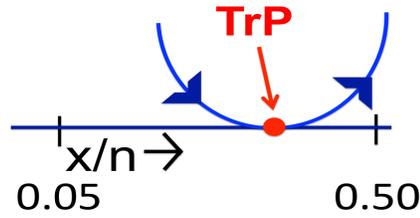

**Figure 2** Path-dependency of transformations of the LR graph shown (A) in terms of the LR itself and (B) projected onto the PV plane by representing each LR by its corresponding (unique) values of P and V. Arrows indicate the direction of "movement" for the given transformation.

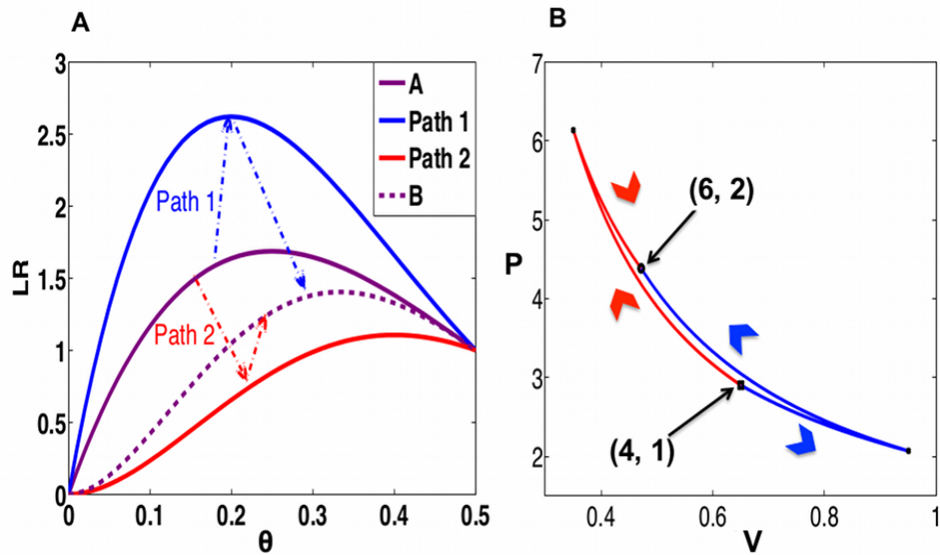

20arXiv September 2013

**Figure 3** Correspondence between the behavior of E and the properties P1-P4 of evidence as shown in Table 1: (A) Behavior of E as a function of x/n for different values of n, displaying properties P1-P2; (B) "isotherms" of the evidential system (holding E constant), displaying property P3; (C) changes in E as a function of increasing n, starting with the (*n*, *x*) pair (1, 0) and increasing by one tail at a time (i.e., going from (1, 0) to (2, 0) to (3, 0)…), illustrating property P4.

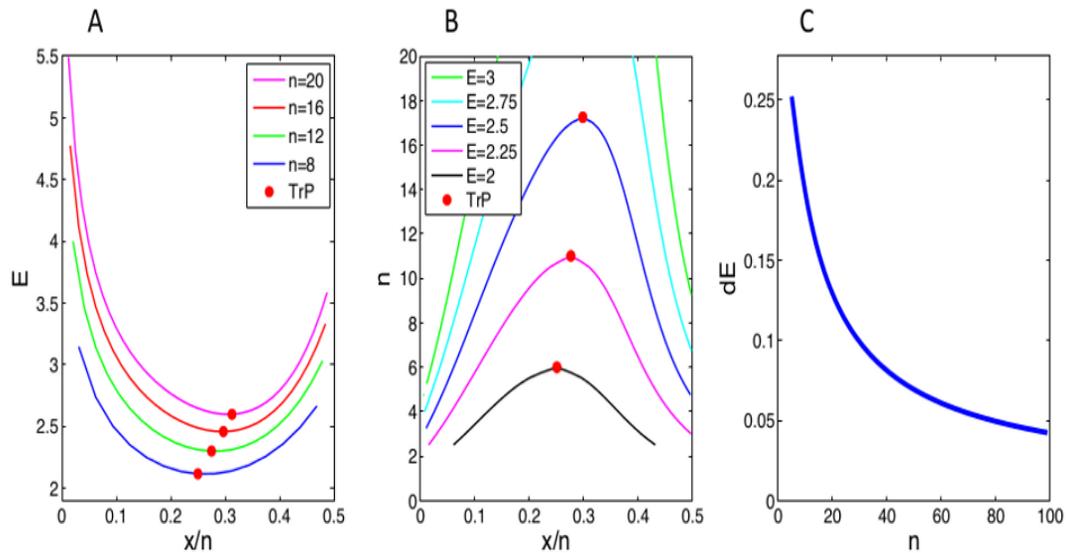